\documentclass{ws-procs9x6}

\def\msbar{\overline{\rm MS}}

\begin{document}

\title{MRST Parton Distributions -- status 2006.}

\author{\vspace{-0.7cm}R.S. Thorne \footnote{\uppercase{R}oyal 
\uppercase{S}ociety \uppercase{U}niversity \uppercase{R}esearch 
\uppercase{F}ellow.}}

\address{\vspace{-0.01in}Department of Physics and Astronomy\\
University College London\\
Gower St, 
London, WC1E 6BT, UK}

\author{\vspace{-0.3cm}A.D. Martin and W.J. Stirling}

\address{\vspace{-0.01in}Institute for Particle Physics Phenomenology\\
University of Durham, DH1 3LE, UK}

\maketitle

\abstracts{\vspace{-0.6cm}We present the new 
preliminary MRST parton distributions at NLO and 
NNLO. The analysis includes some new data and there is an improvement in the 
theoretical treatment at NNLO. Essentially complete NNLO partons 
are presented for the first time, together with uncertainties.}

\vspace{-0.7cm}

There are a number of reasons for an update of the MRST parton distributions. 
Firstly, there are new data to be included: NuTeV data\cite{NuTeVsf} on 
$F_2^{\nu,\bar\nu}(x,Q^2)$ and $F_3^{\nu,\bar\nu}(x,Q^2)$
replacing CCFR\cite{CCFR}; 
new CDFII high-$E_T$ jet data\cite{newjets} (only compared,
not yet fit); and we now include direct high-$x$ data on $F_L(x,Q^2)$.
There are also major changes in the theory: an implementation of 
a new heavy flavour VFNS\cite{nnlovfns}, 
particularly at NNLO; and the inclusion of 
NNLO corrections\cite{NNLODY} to the Drell-Yan cross-sections. 
This leads to some important changes as NLO 
$\to$ NNLO. The most important change compared to the previous 
NNLO partons\cite{MRST04}, which already used the exact splitting 
functions\cite{NNLOs},  
is the   new VFNS which leads to a significant change in the gluon
and heavy quarks. Moreover, due to the NNLO procedure  
being essentially complete we now examine the uncertainties on the NNLO 
partons.  In general the size of the uncertainties 
due to experimental errors is similar to that at NLO\cite{MRSTerror1}. 
There is more work to do in order to estimate the theoretical uncertainty, 
which is 
certainly important in some regions\cite{MRSTerror2}.

We first consider the new data in the fit.
The NuTeV structure function data are not completely compatible with the 
older CCFR data. 
The main source of the discrepancy is in the calibration of the 
magnetic field map of  the
muon spectrometer, i.e. in the muon energy scale.  
However, the previous parton distribution fits were perfectly 
compatible with the CCFR data using an EMC inspired 
$Q^2$-independent nuclear correction\cite{MRST98} $R$. 
This correction is far too large for the 
new NuTeV data. The high-$x$ region is completely
dominated by the valence quarks for both $F_2^{\nu,\bar\nu}(x,Q^2)$ 
and $F_3^{\nu,\bar\nu}(x,Q^2)$.
These are well known from fixed target $F^{\mu,p}_2(x,Q^2)$
and $F^{\mu,d}(x,Q^2)$. In order to fit the NuTeV data we
try a reduced correction factor $R^{eff}=1+A*(R-1)$. 
The best fit is for $A=0.2$ and the previous nuclear correction is 
clearly ruled out. Hence, the NuTeV data imply a nuclear correction 
which is different for neutrinos than for charged leptons. 
However, recent CHORUS\cite{Chorus} data are 
in much better agreement with the  
CCFR data than the NuTeV data. Also, the partons in the region of high 
nuclear correction are already well determined. It may be appropriate to cut 
the nuclear target data in this region. 
The important information that neutrino DIS gives on the flavour 
composition of the proton is in the 
region $x< 0.3$, where the nuclear corrections are not so large or uncertain.

\begin{figure}[ht]
\vspace{-0.6cm}
\centerline{\hspace{-0.9cm}\epsfxsize=2in\epsfbox{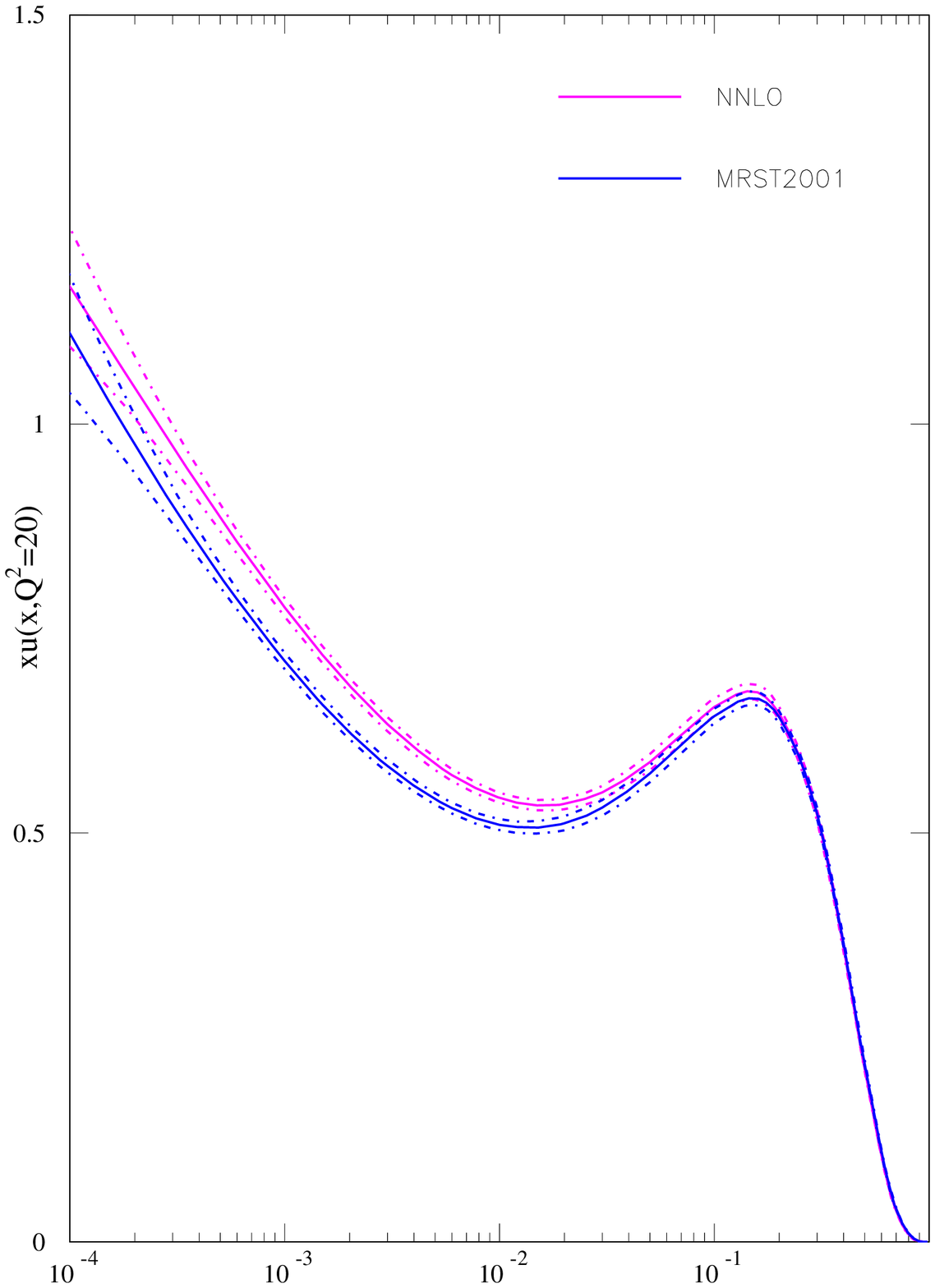}
\hspace{0.5cm}\epsfxsize=2in\epsfbox{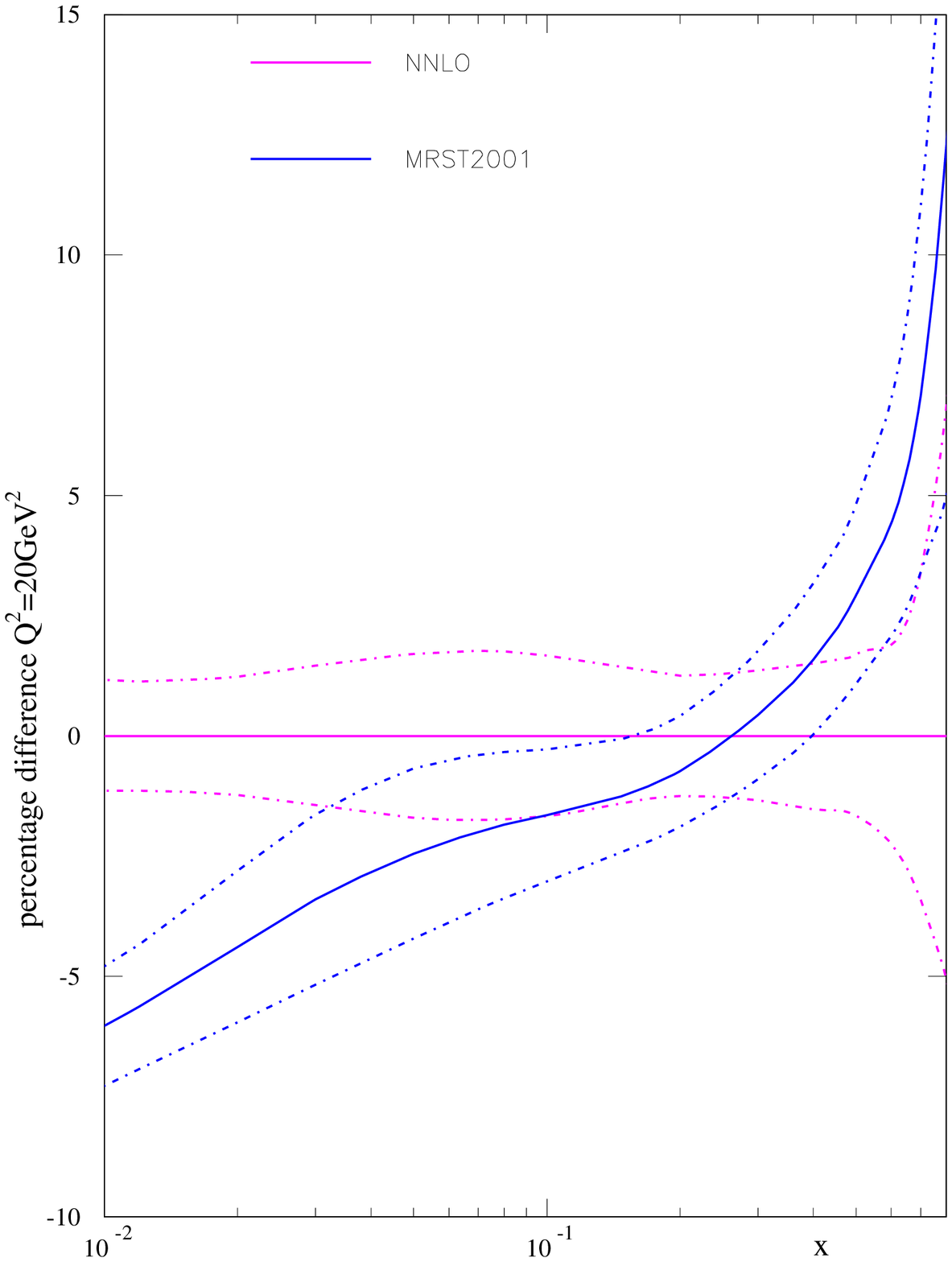}}   
\vspace{-0.1cm}
\caption{Comparison of the NLO up distribution with the NNLO up distribution,
concentrating on small $x$ (left) and high $x$ (right).
  \label{fig1}}
\vspace{-0.5cm}
\end{figure}

The comparison of the most recent MRST partons\cite{MRST04} can be seen 
alongside 
the Tevatron jet data\cite{newjets}. The prediction 
clearly matches quite well, and it 
is clear 
that the new data still require the large high-$x$ gluon in the $\msbar$ 
scheme. The fit to the fixed target data on $F_L(x,Q^2)$ also prefers the 
larger gluon since the data are generally larger than NLO or 
NNLO\cite{MRSTfl}, and a large coupling (and/or higher twist contributions) 
is needed.

The change in the up distribution when going from NLO to NNLO is shown in 
Fig.~\ref{fig1}. At small $x$ the effect of the coefficient functions,
particularly $C_{2,g}(x,Q^2)$, is important and
the difference between  the NLO and the NNLO distribution is 
greater than the uncertainty 
in each calculated using the Hessian approach\cite{CTEQHes}.
At large  $x$ the coefficient functions are again important --
$C^2_{2,q}(x)\sim (\ln^3(1-x)/(1-x))_+$ and the 
difference between NLO and NNLO is again larger than the uncertainty 
in each. There is no real change from the MRST2004NNLO partons for the 
light quarks.
At small $x$ the effect of the splitting functions is important, 
particularly from $P^2_{qg}(x,Q^2)$, which has a
positive $\ln(1/x)/x$ contribution.
This affects the gluon distribution via the fit to $dF_2(x,Q^2)/d\ln Q^2$,
and the NNLO gluon is  smaller at very low $x$ than the NLO gluon.

\begin{figure}[ht]
\vspace{-0.5cm}
\centerline{\hspace{-0.9cm}\epsfxsize=2in\epsfbox{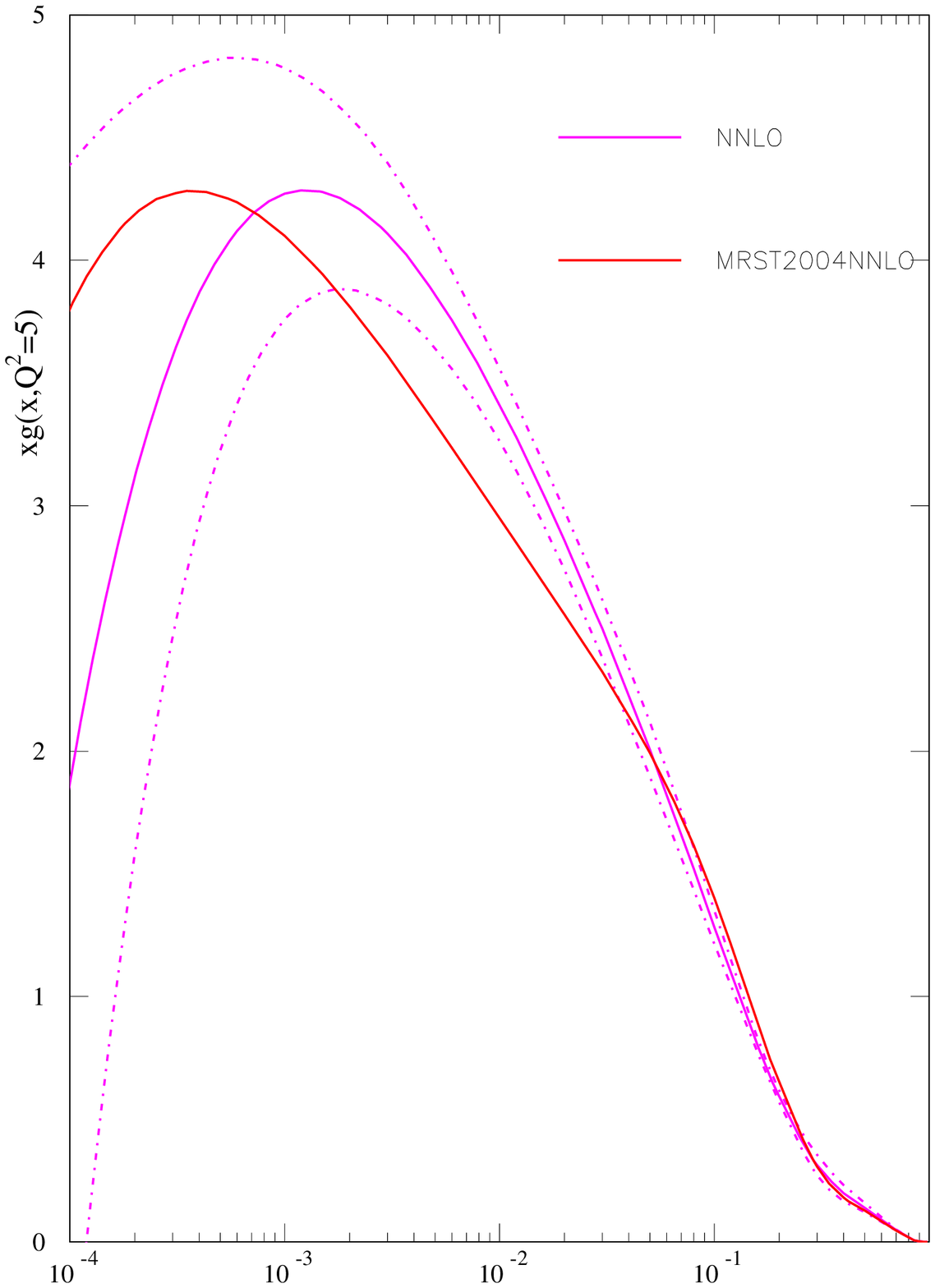}
\hspace{0.5cm}\epsfxsize=2in\epsfbox{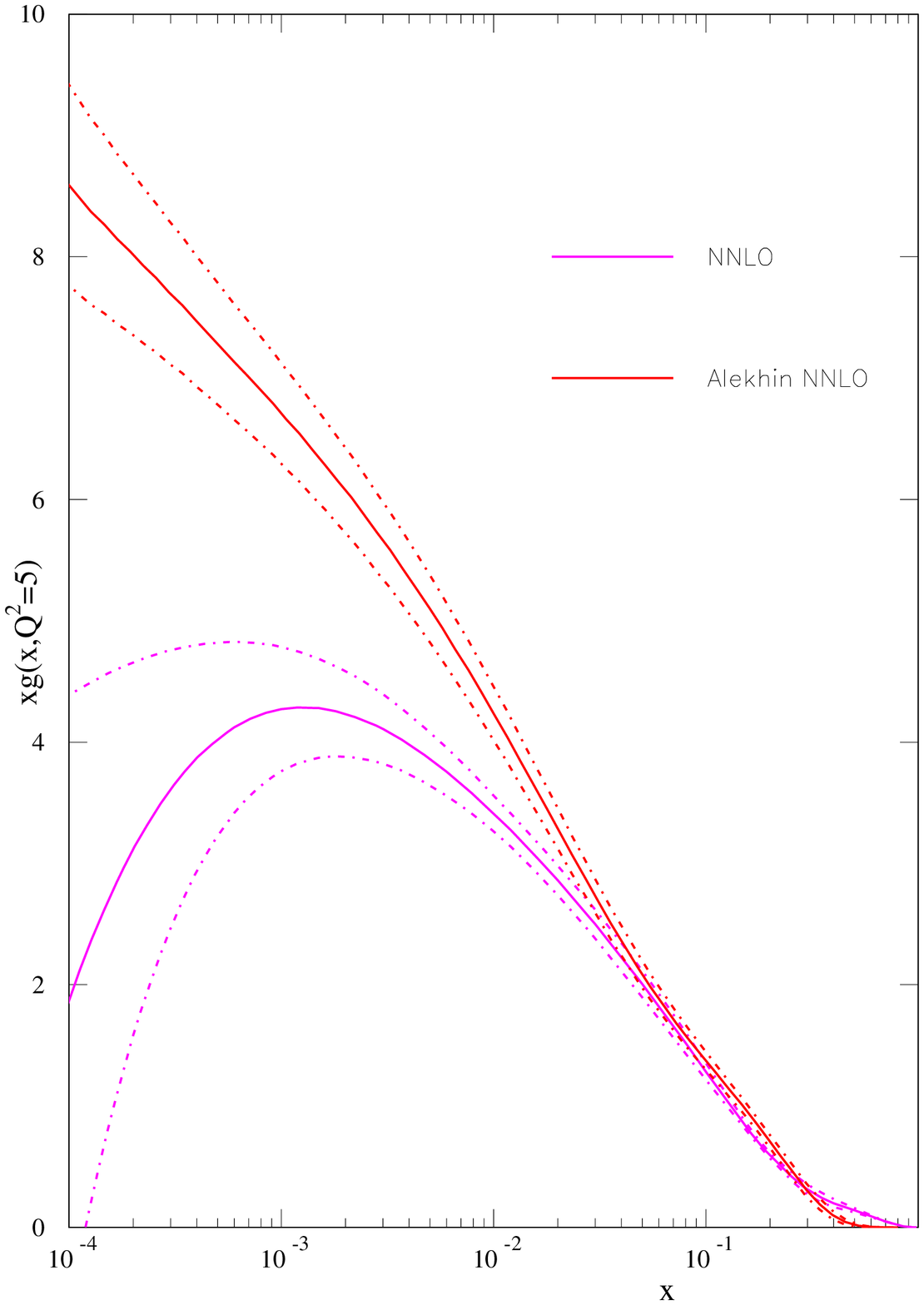}}   
\vspace{-0.1cm}
\caption{Comparison of the NNLO MRST gluon distribution with the previous 
approximate NNLO distribution (left) and
the NNLO distribution of Alekhin (right).
  \label{fig2}}
\vspace{-0.4cm}
\end{figure}

At NNLO heavy flavour no longer evolves from zero at $\mu^2=m_c^2$, i.e. 
$(c+\bar c)(x,m_c^2) = A^2_{Hg}(m_c^2)\otimes g(m_c^2)$.
In practice it starts from a negative value since the matrix element is 
negative at small $x$. The increased evolution from the  
NNLO splitting function allows the charm distribution to catch up partially 
with respect to that at NLO, which starts from zero at $m_c^2$
but it always lags a little at higher $Q^2$. The correct NNLO charm 
is smaller than the approximate  MRST2004
distribution which turned on from zero. This 
correction in the charm procedure also affects the gluon compared to 
the MRST2004 NNLO partons, Fig.~\ref{fig2}, and the  
change is greater than the uncertainty in some places. The correct heavy 
flavour treatment is vital.

\begin{figure}[ht]
\centerline{\hspace{-0.9cm}\epsfxsize=2in\epsfbox{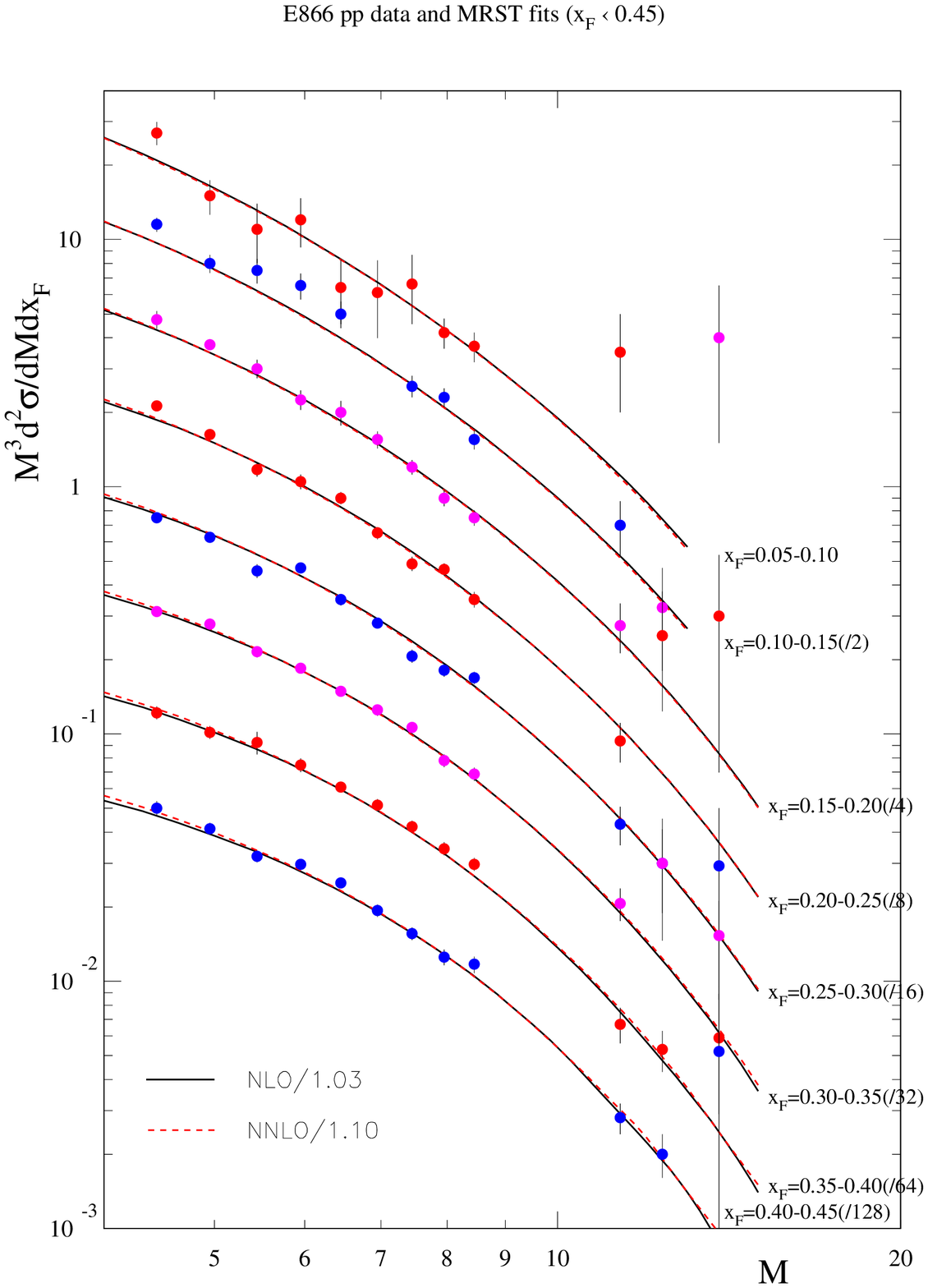}
\hspace{0.5cm}\epsfxsize=2in\epsfbox{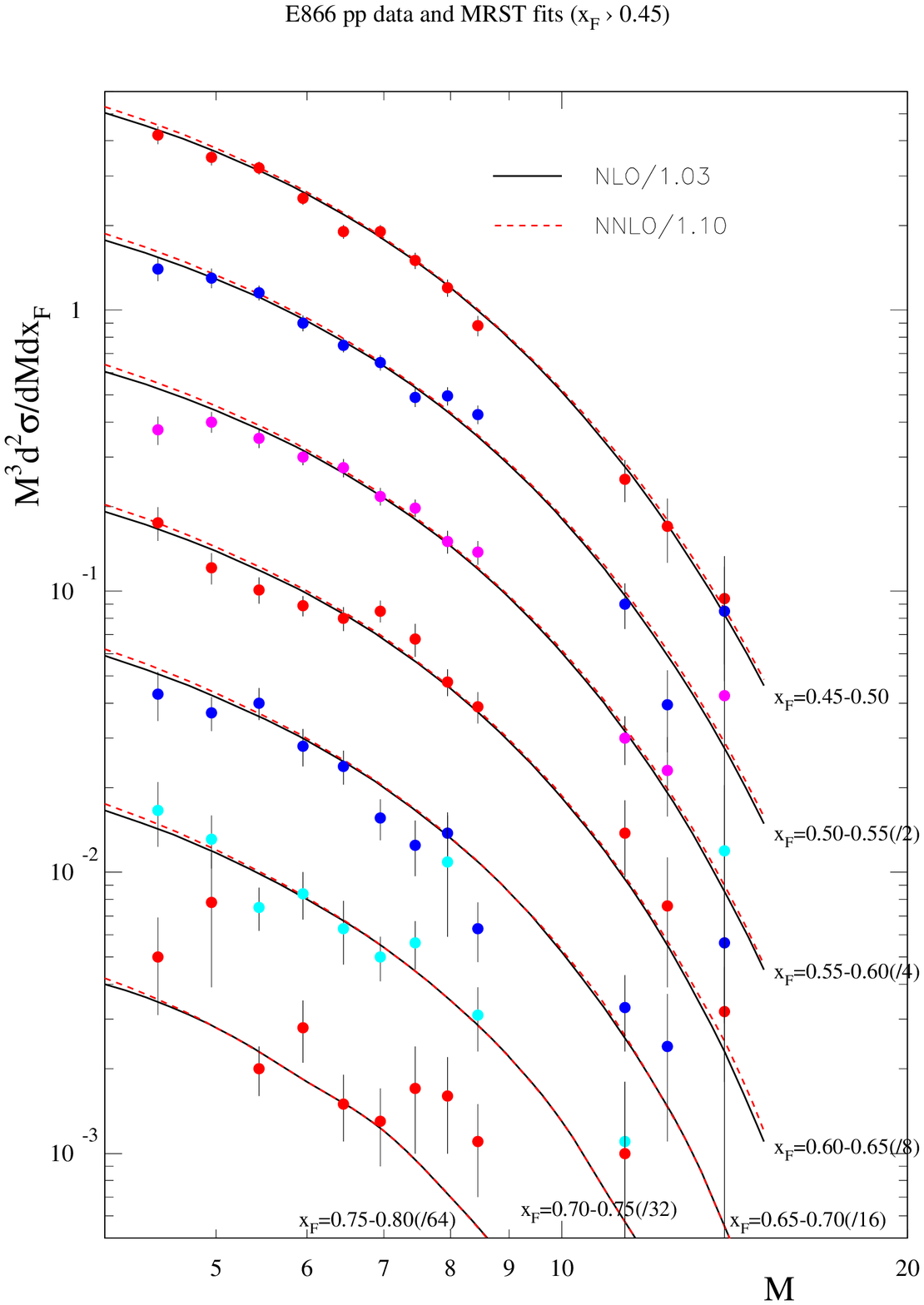}}   
\vspace{-0.2cm}
\caption{Comparison of the NLO MRST Drell-Yan cross-sections with the 
data. 
  \label{fig3}}
\vspace{-0.5cm}
\end{figure}

At NNLO the Drell-Yan corrections\cite{NNLODY} are significant. 
There is an enhancement at high $x_F=x_1\!-\!x_2$ due to large logarithms,
which is similar 
to  the $\ln(1-x)$ enhancement in structure functions. 
The NLO correction is large and the  NNLO corrections are $10\%$ or 
more.  The quality of the fit to E866 Drell-Yan production\cite{E866} 
in proton-proton collisions is $\chi^2= 223/174$ at  NLO and
$\chi^2= 240/174$ at  NNLO. The
scatter of points is large and a $\chi^2 \sim 220$ is the best 
possible. The quality of the fit is good, as seen in Fig.~\ref{fig3}. 
It is worse for proton-deuteron data. The positive 
correction at NNLO requires the data normalization to be $110\%$ 
($103\%$ at NLO), 
there being little freedom since the sea quarks for 
$x\leq 0.1$ and the valence quarks are already 
well determined by structure function data.
The normalization uncertainty on the data is $6.5\%$, and a  change of $10\%$
is a little surprising. 

The quality of the full fit at NLO is $\chi^2=2406/2287$ and at NNLO is
$\chi^2=2366/2287$. NNLO is fairly consistently better than NLO. 
There is a definite tendency for $\alpha_S(M_Z^2)$ to increase with
all changes, both the new data and the improved theoretical treatment. 
At NLO $\alpha_S(M_Z^2)=0.121$ and at NNLO $\alpha_S(M_Z^2)=0.119$.
Although the fit is generally good, particularly at NNLO, there  is 
some room for improvement, and the data would prefer a little more gluon at 
both high and moderate $x$.

We compare with the only other NNLO partons available, those of 
Alekhin\cite{Alekhin} (there is nothing at present from CTEQ). 
We have a much larger $\alpha_S(M_Z^2)$, i.e. 
$\alpha_S(M_Z^2)=0.119$ compared to $0.114$. 
There is not much difference in high-$x$ valence quarks,
except that explained by the difference in $\alpha_S(M_Z^2)$.   
There are differences in the low-$x$ sea quarks but these are
dominated by differences in 
flavour treatments of $\bar u -\bar d$ and $s(x,Q^2)$. 
The gluon distribution difference at small $x$ is seen in Fig.~\ref{fig3}, 
and is much bigger than the uncertainties. This is due to  
the  heavy flavour treatments, which we have already shown to be   
important as well as to differences in the data fit and in $\alpha_S(M_Z^2)$.
The gluons also differ a great deal at high $x$, where they 
are determined by the Tevatron jet data\cite{jets} for MRST, the 
comparison now being excellent\cite{MRST04}.  
In the $\msbar$ scheme the gluon is more important for jets at high 
$x$ at NNLO than at NLO because the high-$x$ quarks are smaller.

Hence, we have included both new data and new theoretical corrections 
in our global analysis. The NNLO fit improves on that at NLO. 
For both the value of $\alpha_S(M_Z^2)$ creeps upwards. 
The NNLO procedure is essentially complete and we have a preliminary update 
of parton distributions. There are more new data to be included -- 
HERA jets\cite{HERAjets}, updated Tevatron high-$E_T$ jets, any
further new heavy flavour data from HERA and a full treatment of 
NuTeV di-muon data.
We also need some further  theoretical fine-tuning, and will 
have fully updated NLO and NNLO partons for 
the LHC complete with uncertainties -- both experimental and theoretical.

\end{document}